\documentclass[%
 aip,jcp,
 amsmath,amssymb,
 reprint,%
]{revtex4-1}

\usepackage{amsmath,amssymb}
\usepackage[scr]{rsfso}
\usepackage{physics}
\usepackage{tensor}
\usepackage{siunitx}
\usepackage{graphicx,subcaption}
\usepackage{chemfig}
\usepackage{bm}
\usepackage{cancel}

\newcommand{\caE}{\mathcal{E}}

\newcommand{\bfc}{\mathbf{c}}

\newcommand{\bfG}{\mathbf{G}}
\newcommand{\bfH}{\mathbf{H}}

\newcommand{\bfnul}{\mathbf{0}}

\newcommand{\bmkappa}{{\bm\kappa}}

\newcommand{\rme}{\mathrm{e}}

\newcommand{\squad}[1]{%
    \newcount\mycount%
    \mycount=0%
    \loop%
        \ifnum\mycount<#1%
            \quad%
            \advance\mycount by 1%
    \repeat%
}

\makeatletter
\let\save@mathaccent\mathaccent
\newcommand*\if@single[3]{%
  \setbox0\hbox{${\mathaccent"0362{#1}}^H$}%
  \setbox2\hbox{${\mathaccent"0362{\kern0pt#1}}^H$}%
  \ifdim\ht0=\ht2 #3\else #2\fi
  }
\newcommand*\rel@kern[1]{\kern#1\dimexpr\macc@kerna}
\newcommand*\wideaccent[2]{\@ifnextchar^{{\wide@accent{#1}{#2}{0}}}{\wide@accent{#1}{#2}{1}}}
\newcommand*\wide@accent[3]{\if@single{#2}{\wide@accent@{#1}{#2}{#3}{1}}{\wide@accent@{#1}{#2}{#3}{2}}}
\newcommand*\wide@accent@[4]{%
  \begingroup
  \def\mathaccent##1##2{%
    \let\mathaccent\save@mathaccent
    \if#42 \let\macc@nucleus\first@char \fi
    \setbox\z@\hbox{$\macc@style{\macc@nucleus}_{}$}%
    \setbox\tw@\hbox{$\macc@style{\macc@nucleus}{}_{}$}%
    \dimen@\wd\tw@
    \advance\dimen@-\wd\z@
    \divide\dimen@ 3
    \@tempdima\wd\tw@
    \advance\@tempdima-\scriptspace
    \divide\@tempdima 10
    \advance\dimen@-\@tempdima
    \ifdim\dimen@>\z@ \dimen@0pt\fi
    \rel@kern{0.6}\kern-\dimen@
    \if#41
      #1{\rel@kern{-0.6}\kern\dimen@\macc@nucleus\rel@kern{0.4}\kern\dimen@}%
      \advance\dimen@0.4\dimexpr\macc@kerna
      \let\final@kern#3%
      \ifdim\dimen@<\z@ \let\final@kern1\fi
      \if\final@kern1 \kern-\dimen@\fi
    \else
      #1{\rel@kern{-0.6}\kern\dimen@#2}%
    \fi
  }%
  \macc@depth\@ne
  \let\math@bgroup\@empty \let\math@egroup\macc@set@skewchar
  \mathsurround\z@ \frozen@everymath{\mathgroup\macc@group\relax}%
  \macc@set@skewchar\relax
  \let\mathaccentV\macc@nested@a
  \if#41
    \macc@nested@a\relax111{#2}%
  \else
    \def\gobble@till@marker##1\endmarker{}%
    \futurelet\first@char\gobble@till@marker#2\endmarker
    \ifcat\noexpand\first@char A\else
      \def\first@char{}%
    \fi
    \macc@nested@a\relax111{\first@char}%
  \fi
  \endgroup
}
\makeatother

\usepackage[english]{babel}
\usepackage[utf8]{inputenc}
\usepackage[T1]{fontenc}

\usepackage{amsmath,amssymb,amsfonts}
\usepackage{mathtools}
\usepackage{mathptmx}
\usepackage{etoolbox}
\usepackage{physics}

\usepackage{gnuplottex}
\usepackage{float}
\usepackage{array}
\usepackage[table,dvipsnames]{xcolor}
\usepackage{graphicx}
\graphicspath{ {./Figures/} }
\usepackage{dsfont}
\usepackage[colorlinks=true, linkcolor=blue, citecolor=blue]{hyperref}
\usepackage{algorithm}
\usepackage[noend]{algpseudocode}
\usepackage{bbold}
\usepackage{bm,bbm}
\usepackage{tabularx, booktabs}
\usepackage{xspace}
\usepackage{listings}
\usepackage{tikz}
\usepackage{bbding}
\usepackage{placeins} % For the FloatBarrier command (to prevent figures from floating into other sections
\usepackage{standalone}
\usetikzlibrary{arrows.meta}
\usetikzlibrary{shapes}
\usepackage{ulem}
\usepackage{dcolumn}% Align table columns on decimal point 
\usepackage{multirow}

\AtBeginEnvironment{bmatrix}{\setlength{\arraycolsep}{2.5pt}}

\makeatletter
\def\mathcolor#1#{\@mathcolor{#1}}
\def\@mathcolor#1#2#3{%
  \protect\leavevmode
  \begingroup
    \color#1{#2}#3%
  \endgroup
}
\makeatother

\newcommand{\be}{\begin{eqnarray}}
\newcommand{\ee}{\end{eqnarray}}
\newcommand{\ocdvtensor}[3]{{#1}_{#2}^\text{#3}}
\makeatletter
\def\@email#1#2{%
 \endgroup
 \patchcmd{\titleblock@produce}
  {\frontmatter@RRAPformat}
  {\frontmatter@RRAPformat{\produce@RRAP{*#1\href{mailto:#2}{#2}}}\frontmatter@RRAPformat}
  {}{}
}%
\makeatother

\begin{document}

\preprint{AIP/123-QED}

\title[]{Orthogonally Constrained CASSCF Framework: Newton-Raphson Orbital Optimization and Nuclear Gradients}
% \title[]{Orthogonally Constrained CASSCF framework for ground and excited states calculation}

\author{Loris Delafosse}
\affiliation{Laboratoire de Chimie Quantique, Institut de Chimie,
CNRS/Université de Strasbourg, 4 rue Blaise Pascal, 67000 Strasbourg, France}

\author{Vincent Robert}
\affiliation{Laboratoire de Chimie Quantique, Institut de Chimie,
CNRS/Université de Strasbourg, 4 rue Blaise Pascal, 67000 Strasbourg, France}

\author{Saad Yalouz}
\email{saad.yalouz@cnrs.fr}
\affiliation{Laboratoire de Chimie Quantique, Institut de Chimie,
CNRS/Université de Strasbourg, 4 rue Blaise Pascal, 67000 Strasbourg, France}

\date{\today}

\begin{abstract}

In a recent work~\cite{yalouz2023ocoo}, we introduced the foundations of an orthogonally constrained complete active space self-consistent field (OC-CASSCF) framework that produces state-specific molecular orbitals for mutually orthogonal  multiconfigurational electronic states. 
In the present study, we extend this approach by incorporating a Newton-Raphson orbital-optimization scheme, for which we derive analytical expressions of the orbital gradient and Hessian. 
Furthermore, we outline a practical route toward the evaluation of analytical nuclear gradients, enabling geometry optimizations within the OC-CASSCF formalism. 
Benchmark calculations on the three lowest singlet states of LiH and H$_2$O molecules demonstrate a systematic improvement as compared to conventional state-averaged CASSCF, even when using modestly sized active spaces.
 
\end{abstract}

\maketitle

\section*{Introduction}
\label{sec:introduction}

A central challenge in electronic structure theory concerns the development of computational methods that can accurately describe both ground- and excited-state properties.
This question naturally arises from the increasing need to model complex molecular processes in which strong electronic correlations simultaneously affect multiple states. 
Illustrative examples include photochemical reactions 
{and their applications in therapy} 
~\cite{bonnet2023ruthenium,smith2013recent}, molecular magnetism~\cite{roseiro2025interplay, roseiro2023modifications, sheng1994magnetism, verot2012importance, vela2017electron}, and electron transfer processes~\cite{marcus1993electron,domingo2015electronic}.

In wavefunction-based methodologies, a broad range of approaches has been proposed to address this challenge.
At the single-particle level, extended mean-field methods~\cite{hunt1969orthogonality,gilbert2008self,barca2018simple,thom2008locating} have been developed to optimize molecular orbitals (MOs) for specific excited states.
Beyond mean-field approximations, numerous multiconfigurational methods have been introduced to capture electronic correlation effects.
These include configuration-interaction (CI) expansions combined with Monte-Carlo techniques~\cite{dash2019excited,dash2021tailoring,cuzzocrea2022reference,shepard2022double,pathak2021excited}, density-matrix renormalization group methods~\cite{white1993density,stoudenmire2012studying,chan2016matrix,fishman2022itensor}, coupled-cluster approaches~\cite{marie2021variational,kossoski2021excited,damour2024state}, and even strategies designed for quantum computers~\cite{zhu2025statespecificoo,yalouz2022analytical,Yalouz_2021,higgott2019variational,nakanishi2019subspace}. 
Apart from Full-CI (FCI) expansions, the accuracy of truncated multi-configurational methods 
%is known to depend critically 
depends
on the choice of the MO basis set. 
Optimizing orbitals for specific states therefore emerges as a crucial step. 

Building on this principle,  the Complete-Active-Space Self-Consistent Field (CASSCF) framework~\cite{siegbahn1981complete,malmqvist2002restricted,dobrautz2021spin,manni2023openmolcas} remains a reference. 
% for describing both ground and excited electronic states.
By combining a reduced CAS-CI expansion with orbital optimization, the CASSCF method provides a balanced and variationally motivated 
treatment of static correlation effects.
In its state-averaged formulation (SA-CASSCF), a common set of MOs is optimized over an ensemble of states, yielding mutually orthogonal many-electron wavefunctions. %that share a common \textbf{set of MOs}. 
While ``state averaging'' offers a convenient mathematical trick for multi-state calculations, the use of shared orbitals suppresses state-specific orbital relaxation, making excitation energies and properties sensitive to the number and nature of states included in the ensemble.
% By contrast, the state-specific formulation (SS-CASSCF) allows full orbital relaxation for a targeted state.
% \vincent{Cette fin de phrase est difficile à comprendre. A moins de donner les références 36 et 37. Je mentionnerais les choses ainsi :
% \textbf{By contrast, the state-specific formulation (SS-CASSCF) allows full orbital relaxation for a targeted state, some issues emerge.
% It is well-known that SS-CASSCF....}
By contrast, the state-specific formulation (SS-CASSCF) allows full orbital relaxation for a targeted state. 
However, because it does not satisfy a variational principle for excited states, SS-CASSCF is known to be prone to root flipping and variational collapse toward lower-lying states.
Despite this limitation, note that recent studies have shown that SS-CASSCF can sometimes achieve improved accuracy with smaller CAS~\cite{marie2023excited,saade2024excited} compared to state-averaged approaches.

The variational collapse inherent to SS-CASSCF can be seen as a consequence of the absence of explicit orthogonality constraints between independently optimized states. 
To address this issue, we have recently laid the foundations of an orthogonally constrained (OC) version of CASSCF (OC‑CASSCF)~\cite{yalouz2023ocoo}, in which excited states are optimized to remain orthogonal to previously computed states (note that similar strategies have also been considered in other frameworks~\cite{zhu2025statespecificoo, shepard2022double, pathak2021excited, stoudenmire2012studying, fishman2022itensor,levi2020variational}).
In this letter, we present two main extensions of the OC-CASSCF framework that enhance the method’s applicability. First, to go beyond the brute-force numerical orbital optimization of our previous work~\cite{yalouz2023ocoo}, we introduce a second-order Newton-Raphson scheme that incorporates the orthogonality constraints between multiple states. 
Second, motivated by the need to access response properties with practical applications (e.g. quantum dynamics, geometry optimizations), we propose a route to compute analytical nuclear gradients.
% within the OC-CASSCF method.

% \vincent{Pour l'énconomie de mots, on peut supprimer cette phrase liminaire : The letter...}
The letter is structured as follows. 
In Sec.~\ref{sec:theory}, we introduce the theoretical framework of OC-CASSCF. 
We then discuss practical two-step implementation and provide analytical expressions for the OC orbital Gradients and Hessians required in the Newton-Raphson optimization.
% We also show how a simplified form of the nuclear gradient can be obtained for OC-CASSCF energies.
An analytical form of the nuclear gradients is also presented for the OC-CASSCF energies.
% \vincent{Je propose plutôt que la phrase précédente : 
% A simplified form of the nuclear gradient is derived for the OC-CASSCF energies.}
In Sec.~\ref{sec:numerical}, we illustrate the performance of the method to describe the three lowest singlet states of the LiH molecule (and H$_2$O in Supplementary material \ref{supp:H2O}). 
% The results demonstrate that OC-CASSCF successfully recovers the Full-CI ground, first and second excited states of the two systems, whereas SA-CASSCF fails to describe the dissociation limit (even when applied only to the ground and first excited states).
Finally, the use of the OC-CASSCF gradients is illustrated by performing geometry optimizations of the two molecular excited states.
% \loris{Discuter théorie des perturbations}

% \vincent{Introduction parfaite !}

\section{Theory}
\label{sec:theory}

\subsection{Orthogonally Constrained-CASSCF (OC-CASSCF)
% \vincent{(OC-CASSCF)} framework  
}
\label{subsec:occasscf}

% Within the OC-CASSCF framework, the goal is to approximate a set of low-lying multiconfigurational electronic eigenstates (including both ground and excited states) by optimizing a series of CASSCF wavefunction ans\"{a}tze at both the orbital and  CI levels, while rigorously enforcing mutual orthogonality among all resulting states.

Within the OC-CASSCF method, the goal is to approximate the low-lying eigenspectrum of a many-electron system (ground and excited states)  \textit{via} the optimization, at both orbital and  CI levels, of a series of wavefunction ans\"{a}tze  $ \Bqty{  \ket{\Psi_I (\bmkappa,\bfc)} }  $ while simultaneously enforcing their mutual orthogonality (\textit{i.e.} $  \braket{\Psi_I (\bmkappa,\bfc)}{\Psi_{I'} (\bmkappa',\bfc')} = \delta_{II'} $).   
To proceed, each state is described by a standard CASSCF-like ansatz
\begin{equation}\label{eq:MCSCF_wavefunction}
    \ket{\Psi_I  (\bmkappa,\bfc)} = \rme^{-\hat{\kappa}} \sum_i c_{i} \ket{\Phi_i^\textrm{CAS}},
\end{equation}
with CI coefficients $\mathbf{c}= (\ldots, c_i, \ldots)$ (where  $  \ket{\Phi_i^\textrm{CAS}}$ are active-space Slater determinants or configuration state functions), and MO-rotation parameters $\boldsymbol{\kappa} = (\ldots,\kappa_{pq}, \ldots)$ defining the generatrix $\hat{\kappa} = \sum_{p>q}^\text{MOs} \kappa_{pq}(\hat{E}_{pq}-\hat{E}_{qp})$ 
where $\hat{E}_{pq}=\sum_{\sigma\in{(\uparrow,\downarrow)} }\hat{a}^\dagger_{p,\sigma}\hat{a}^\dagger_{q,\sigma} $ and $\hat{a}^\dagger_{p,\sigma}/\hat{a}_{p,\sigma}$ are spin-orbital creation/annihilation operators with spatial 
and spin parts $p$ and  $\sigma$ respectively.
To optimize these wavefunction ansätze and ensure  mutual orthogonality between the OC-CASSCF states, one proceeds in an incremental way.
First, the MOs and CI coefficients of a given wavefunction ansatz
% \vincent{Toujours pour réduire, je propose de supprimer la parenthèse qui suit}
% (as given by Eq.~(\ref{eq:MCSCF_wavefunction})) 
are optimized to describe a ground state.
Then, to represent the first excited state, a new ansatz (with new CI and orbital parameters) is optimized now including an orthogonality constraint
to enforce zero overlap with the previously optimized ground-state ansatz.
% \saad{J'en suis la !}
% \vincent{Question : l'orthogonalité est-elle
% imposée avec l'ansatz de l'état fondamental, ou avec l'état fondamental optimisé ? La phrase qui suit pour les autres états laisse penser que l'on travaille avec les $K-1$ états optimisés. Ce que je pense...}
This procedure is then systematically extended to higher-lying excited states, such that the $K$-th excited state is orthogonal to the $K$ previously optimized ansatz states $\Bqty{\ket{\Psi_I}}_{I=0}^{K-1}$.
Importantly, at the end of this optimization protocol, each state is characterized by its own independent set of optimal MOs and CI coefficients.

From a formal point of view, optimizing a given $K$-th excited state with the OC-CASSCF scheme can be seen as a combined \textit{``CI+Orbital''} optimization problem over the penalized energy functional $ E_K^{\textrm{OC}} (\bmkappa,\bfc)$ given by
\begin{equation}\label{eq:costfunction}
\begin{split}
    \min_{\bmkappa,\bfc} E_K^{\textrm{OC}} (\bmkappa,\bfc) &= \min_{\bmkappa,\bfc}  \left[ \langle \Psi_K (\bmkappa,\bfc) | \hat{H}^\textrm{OC}_K   | \Psi_K (\bmkappa,\bfc)\rangle \right]  \\
    &= \min_{\bmkappa,\bfc}  \left[ \langle \Psi_K (\bmkappa,\bfc) | (\hat{H}  + \hat{P}_K^\textrm{OC} ) | \Psi_K (\bmkappa,\bfc)\rangle \right]  \\
    &= \min_{\bmkappa,\bfc} \left[  \langle \hat{H}\rangle_{ \Psi_K }  + \langle \hat{P}_K^\textrm{OC}  \rangle_{ \Psi_K  }  \right].
\end{split}
\end{equation}
which satisfies orbital and CI stationarity conditions:
\begin{equation}\label{eq:stationarity}
    \pdv{E_K^{\textrm{OC}}}{\bmkappa} = \bfnul  \qcomma   \pdv{E_K^{\textrm{OC}}}{\bfc} = \bfnul .
\end{equation} 
% \vincent{Si tu manques de place, on peut racourcir ce qui suit ainsi :
The penalized energy functional $E_K^{\textrm{OC}}$ in Eq.~(\ref{eq:costfunction}) is expressed as the expectation value of an ``Orthogonally Constrained (OC) Hamiltonian'' noted ${\hat{H}^\textrm{OC}_K}$.
The first contribution $\langle \hat{H}\rangle_{ \Psi_K} $ is the conventional electronic energy given by the electronic structure Hamiltonian
\begin{equation}
    \hat{{H}} = \sum_{pq} h_{pq} \hat{E}_{pq} + \dfrac{1}{2} \sum_{pqrs} g_{pqrs} \hat{e}_{pqrs} + \caE_\text{Nuc},
\end{equation}
where $h_{pq}$ and $g_{pqrs}$ are the one- and two-electron integrals associated to their respective excitation operators $\hat{E}_{pq}$, and $\hat{e}_{pqrs}=\sum_{pqrs}\hat{E}_{pq}\hat{E}_{rs}-\delta_{qr}\hat{E}_{ps}$
 and $\caE_\text{Nuc}$ is the nuclear repulsion energy.

% }
% As seen in Eq.~(\ref{eq:costfunction}), the penalized energy functional $E_K^{\textrm{OC}}$ is expressed as the expectation value of an ``orthogonally constrained Hamiltonian'', denoted $\hat{H}_K^{\textrm{OC}} = \hat{H} + \hat{P}_K^{\textrm{OC}}$, which gives rise to two contributions.
% The first one, noted $\langle \hat{H}\rangle_{ \Psi_K} $, is the conventional electronic energy, given by the expectation value of the electronic structure Hamiltonian
% \begin{equation}
%     \hat{{H}} = \sum_{pq} h_{pq} \hat{E}_{pq} + \dfrac{1}{2} \sum_{pqrs} g_{pqrs} \hat{e}_{pqrs} + \caE_\text{Nuc},
% \end{equation}
% where $h_{pq}$ and $g_{pqrs}$ are the one- and two-electron integrals, $\hat{E}_{pq}$ and $\hat{e}_{pqrs}=\sum_{pqrs}\hat{E}_{pq}\hat{E}_{rs}-\delta_{qr}\hat{E}_{ps}$ are the one- and two-electron excitation operators, and $\caE_\text{Nuc}$ 
% \vincent{Je croyais que $\caE_\text{Nuc}$ c'était un
% homme dépourvu de ses attributs...JPP.}
%is the nuclear potential energy. 
The second contribution in  Eq.~(\ref{eq:costfunction}), noted $\langle \hat{P}_{K}^\textrm{OC}\rangle_{ \Psi_K} $, enforces orthogonality with respect to previously optimized states. It is defined through the projector  
\begin{equation}\label{eq:OC_proj}
    \hat{P}_K^\textrm{OC}  =  \sum_{I=0}^{K-1} \Delta_I \ketbra{\Psi_I}{\Psi_I}
\end{equation} 
where $\Delta_I$ are the so-called penalty shifts (positive amplitudes) associated to the previously optimized states $\Bqty{\ket{\Psi_I}}_{I=0}^{K-1}$.
Within this framework, any trial state exhibiting a non zero overlap with previously optimized states 
$\{\,|\Psi_I\rangle\,\}_{I=0}^{K-1}$ incurs an energetic penalty proportional to $\Delta_I$, thereby enforcing orthogonality in its optimization.

Interestingly, following the procedure defined in Eq.~(\ref{eq:costfunction}) provides a natural extension of the variational principle to excited states. 
Within this framework, each newly optimized state is constrained to lie above the previously obtained ones in energy, as the imposed orthogonality conditions prevent collapse onto lower-lying solutions (see Refs~\cite{zhu2025statespecificoo, shepard2022double, pathak2021excited, stoudenmire2012studying, fishman2022itensor}).  

\subsection{Practical Implementation: Two-Step Approach }
\label{subsec:two_step}

% Starting from this, targeting the $K$-th low-lying eigenstate within the OC-CASSCF framework involves following an incremental optimization strategy. First, the orbitals and CI coefficients of a given CAS ansatz as defined in Eq.~(\ref{eq:MCSCF_wavefunction}) are optimized to describe the ground state.
% Then, a new CAS ansatz (with new CI and orbital parameters) is optimized to represent the first excited state, now including orthogonality constraints that enforce zero overlap with the previously optimized ground-state ansatz.
% This procedure is then systematically extended to higher-lying excited states, such that the $K$-th state obtained is orthogonal to the $K-1$ previously determined states.
% It is important to note that, at the end of this protocol, each state is characterized by its own independently optimized set of CI coefficients and orbitals. 

% We will now discuss a practical strategy for implementing the OC-CASSCF framework .
As in conventional CASSCF methods, a simultaneous optimization of both CI and orbital parameters is computationally demanding. 
To address this, we build upon our previous work~\cite{yalouz2023ocoo} and adopt a practical ``two-step'' approach that enforces the orthogonality constraints at both the CI and orbital levels. 
Within this framework, the $K$-th electronic state is obtained by decoupling the optimization into two successive steps described below.
% First, the CI coefficients are optimized while keeping the orbitals fixed. 
% Subsequently, the orbitals are optimized while holding the CI parameters fixed.

\paragraph{Orthogonally Constrained Configuration-Interaction:} 
In a first step, the CI parameters $\bfc$ of the state $ \ket{\Psi_K}$ are evaluated by solving the following %orthogonally constrained 
% \vincent{OC} 
OC-CASCI eigenvalue equation (assuming fixed parameters $\bmkappa = \bmkappa^*$):
\begin{equation}\label{eq:CI_prob}
    \hat{P}^{AS}(\bmkappa^*)   \hat{H}^\textrm{OC}_K   \hat{P}^{AS}(\bmkappa^*) \, \ket{\Psi_K} = E_K   \ket{\Psi_K},
\end{equation} 
where $\hat{P}^{AS}(\bmkappa^*)$ projects onto the active-space many-body basis
% \vincent{je pense plutôt que les many-body sont "built on" que "corresponding to"}
% corresponding to 
{built on}
the frozen orbitals $\bmkappa^*$. 
Despite the presence of the subscript ``$K$'', note that Eq.~(\ref{eq:CI_prob}) actually represents a ground-state eigenvalue problem: the penalty terms introduced in Eq.~(\ref{eq:OC_proj}) ensure that the ground state of the projected OC  
Hamiltonian $\hat{P}^{AS}(\bmkappa^*)   \hat{H}^\textrm{OC}_K   \hat{P}^{AS}(\bmkappa^*) $ corresponds to the targeted $K$-th excited state $\ket{\Psi_K}$.
% \vincent{"This is due..." C'est un acte volontaire ! Je tourne
% la phrase un peu différemment. Et j'aimerais être certain de ne pas
% avoir mal transcrit. Autre point même si je milite pour l'explicitation, en évitant les acronymes. Ici, il serait quasiment le seul. orthogonally constrained = OC.
% Donc par exemple "projected orthogonally constrained Hamiltonian" deviendrait "projected OC Hamiltonian". Et de faire la chasse ensuite à tous les "orthogonally constrained" du texte. Ce matin, je suis revenu en amont en introduisant l'acronyme "OC". En préservant l'explicitation
% en commentaire, des fois que l'acronyme vous paraitrait ajouter de la lourdeur. Je lis parfois : QPE-LMPDFT-FullCI-SC method...! J'ai ajouté un trait d'union partout 
% $\rightarrow$ orthogonally constrained, pour homogénéiser. Sauf oubli, évidemment.}
% This is due to the penalty terms introduced in Eq.~(\ref{eq:OC_proj}), which ensure that the effective ground state of this projected Hamiltonian corresponds to the excited state $\ket{\Psi_K}$.% (rather than the literal $K$-th state of the active-space problem).

\paragraph{Orthogonally Constrained Orbital-Optimization:} 
The second step consists in optimizing the orbital parameters $\bmkappa$ with fixed CI coefficients $\bfc^\ast$, while still enforcing orthogonality constraints. 
In our previous work~\cite{yalouz2023ocoo}, this orbital optimization was carried out in a ``brute-force'' manner using numerical optimizers, which becomes impractical for larger-scale systems. 
Here, we introduce a more efficient and systematic strategy based on a 
% \vincent{par rapport à la référence 1, l'approche ici est perturbative. Je ne suis pas très au fait des usages, mais j'insisterais sur cet aspect, contrastant avec le numérique mais "exact". $\rightarrow$ "perturbatively-driven"}
second-order Newton-Raphson algorithm~\cite{helgaker2014molecular} incorporating orthogonality constraints.
Within this framework, the orbital update is given by
\begin{equation}\label{eq:NR_step}
     \bmkappa \leftarrow \bmkappa - \pqty{\bfH^\textrm{OC}}^{-1} \bfG^\textrm{OC}
\end{equation}
where we have introduced the 
%orthogonally constrained 
{OC}
orbital Gradient $\bfG^\textrm{OC}$ and Hessian $\bfH^\textrm{OC}$ including both an energy and an overlap part:
\begin{gather}
\begin{split}
    G_{pq}^\text{OC}    &=    G_{pq}^\text{Energy} + G_{pq}^\text{Overlap}, \\
    H_{pq,rs}^\text{OC} &= H_{pq,rs}^\text{Energy} + H_{pq,rs}^\text{Overlap}.
\end{split}
\end{gather}
Analytical expressions for the energy contributions to the Gradient and Hessian are already well-known in the literature (see Refs~\cite{helgaker2014molecular, yalouz2022analytical} for more details).
% \vincent{On comprend que ces éléments sont "well-known". Mais la phrase qui suit est troublante puisque le Hessian est "to our knowledge"...Pour les overlaps, je comprends l'originalité, mais cette dernière ne s'applique pas pour le Hessian. J'imagine que tu veux faire référence au caractère "OC" ?}
Here, we present the analytical forms of the overlap contributions, which, to our knowledge, have never been introduced (see Supplementary Material~\ref{supp:orbital_opt} for a detailed derivation and possible simplifications):
\begin{subequations}\label{eq:orbital_derivatives}
\begin{gather}
    \label{subeq:orbital_gradient}
    \ocdvtensor{G}{pq}{Overlap} = 2 \sum_{I=0}^{K-1} \Delta_I S^{\Psi_K,\Psi_I} ( \gamma^{\Psi_K,\Psi_I}_{pq} - \gamma^{\Psi_K,\Psi_I}_{qp} ), \\
    \begin{split}\label{subeq:orbital_hessian}
    &\ocdvtensor{H}{pq,rs}{Overlap} = 2 \sum_{I=0}^{K-1} \Delta_I(\gamma^{\Psi_K,\Psi_I}_{pq}-\gamma^{\Psi_K,\Psi_I}_{qp}  )(\gamma^{\Psi_K,\Psi_I}_{rs}-\gamma^{\Psi_K,\Psi_I}_{sr})
    \\
    &+ \sum_{I=0}^{K-1} \Delta_I S^{\Psi_K,\Psi_I} \bigg( 2 (\Gamma^{\Psi_K,\Psi_I}_{pqrs} - \Gamma^{\Psi_K,\Psi_I}_{qprs} - \Gamma^{\Psi_K,\Psi_I}_{pqsr} + \Gamma^{\Psi_K,\Psi_I}_{qpsr}) \\
    &+ \delta_{rq} (\gamma^{\Psi_K,\Psi_I}_{ps} + \gamma^{\Psi_K,\Psi_I}_{sp}) 
    - \delta_{sq}(\gamma^{\Psi_K,\Psi_I}_{p r} + \gamma^{\Psi_K,\Psi_I}_{rp}) \\
    &- \delta_{rp}(\gamma^{\Psi_K,\Psi_I}_{qs} + \gamma^{\Psi_K,\Psi_I}_{sq} ) 
    + \delta_{sp}(\gamma^{\Psi_K,\Psi_I}_{qr} + \gamma^{\Psi_K,\Psi_I}_{rq} )
    \bigg)
    \end{split}
\end{gather}
\end{subequations}
%\vincent{la formule (9b) ci-dessus bave un peu à droite.} 
where $S^{\Psi_K,\Psi_I} = \braket{\Psi_K}{\Psi_I} $, while $\gamma^{\Psi_K,\Psi_I}_{pq} = \mel{\Psi_K}{\hat{E}_{pq}}{\Psi_I}$ and $\Gamma^{\Psi_K,\Psi_I}_{pqrs}   = \mel{\Psi_K}{\hat{e}_{pqrs}}{\Psi_I}$ represent one- and two-electron transition reduced density matrices (TRDMs) respectively. 
These quantities are discussed in Supplementary Material~\ref{supp:cas_trdm}.

Implementing the OC Newton-Raphson step as defined in Eq.~(\ref{eq:NR_step}) provides a simultaneous ``energy + overlap'' optimization for the targeted state $\Psi_K$ at the orbital level. 
At the end of this orbital optimization, the updated MOs are used to build a new CAS-CI problem to be solved, as given in Eq.~(\ref{eq:CI_prob}). 
This alternating procedure between CI and orbital optimization is iterated self-consistently until convergence in the total energy.

Importantly, while we focus here on a two-step approach, an alternative “one-step” implementation of the OC-CASSCF method is also possible in principle. It would consist in a generalized Newton-Raphson optimization that simultaneously updates both the CI and orbital parameters~\cite{helgaker2014molecular}. In addition to the orbital derivatives of Eq.~(\ref{eq:orbital_derivatives}), this scheme would require the OC Gradient and Hessian with respect to the CI coefficients, as well as the mixed CI-orbital Hessian. Although not the main focus of this work, analytical expressions of these quantities are provided in Supplementary Material~\ref{supp:configuration_opt} for the sake of completeness.

\subsection{Analytical OC-CASSCF Nuclear Gradients}
\label{subsec:nuc_grad}

% To conclude,
% \vincent{C'est un peu tôt pour conclure, non ?! Je fais
% une autre proposition juste après.}
% we want to outline a route to compute OC-CASSCF nuclear gradients for both ground and excited states.

% \vincent{As a follow up of OC electronic states spectrum 
% construction, we felt that the strategy could be extended to compute OC-CASSCF nuclear gradients for both ground and excited states.}

As a follow-up on OC-CASSCF electronic states and spectrum construction, we also introduce a strategy to compute the nuclear gradients for both ground and excited states.
From a practical perspective, these gradients are essential for applications such as geometry optimization and the computation of nuclear forces in \textit{ab initio} quantum dynamics. 
Accessing an analytical form of such gradients is particularly important, as it offers significant advantages in numerical stability and computational efficiency compared to purely numerical finite-difference approaches.

To derive an analytical expression for the OC-CASSCF gradients, we build on the framework discussed in Refs.~\cite{helgaker1984second,helgaker1998gradient,yamaguchi2011analytic}.
More specifically, in fully variational state-specific electronic structure methods where both orbital and CI parameters are fully optimized (such as ground-state CASSCF or Hartree-Fock), 
%it has been shown that 
% \vincent{petite reformulation}
the gradient of any individual electronic state 
can be substantially simplified. 
% In particular, the total derivative of the energy with respect to a nuclear coordinate $x$ reduces
% \vincent{C'est curieux. En effet, la réduction apparait à la fin de l'équation 10. Donc je donnerais : 1. l'expression, puis 2. l'annulation des gradients de l'énergie, pour dans un dernier temps 3. donner l'égalité simplifiée. Je fais une proposition ci-dessous, en maintenant le texte actuel}
% to
% \begin{equation}\label{eq:simplified_analytical_form}
% \begin{split}
%     \dv{E}{x} &= \pdv{E}{x}
%     + \pdv{E}{\bfc} \pdv{\bfc}{x}
%     + \pdv{E}{\bmkappa} \pdv{\bmkappa}{x} \\
%     &= \pdv{E}{x}
%     = \bra{\Psi} \pdv{\hat{H}}{x} \ket{\Psi}.
% \end{split}
% \end{equation}
% This result follows directly from the CI and Orbital stationarity conditions
% $\partial E / \partial \bfc = 0 $ and $\partial E / \partial \bmkappa = 0$ fulfilled within such fully variational methods which reduce the nuclear derivative to the expectation value of the Hamiltonian derivative operator $\partial \hat{H} / \partial x$ (see Refs.~\cite{helgaker1984second,helgaker1998gradient,staalring2001analytical,yalouz2022analytical} and Supplementary Material~\ref{supp:nuclear_gradient}).
% % \vincent{SUGGESTION à la place du paragraphe précédent ci-après, en découpant l'équation en 2. J'ai mis les labels ....form1 et ....form2
% % qui donnent 11 et 12, et donneront donc 10 et 11 si vous êtes ok.} 
In particular, the total derivative of the energy $E$ of a given state with respect to a nuclear coordinate $x$ reads
\begin{equation}\label{eq:simplified_analytical_form_1}
\begin{split}
    \dv{E}{x} &= \pdv{E}{x}
    + \pdv{E}{\bfc} \pdv{\bfc}{x}
    + \pdv{E}{\bmkappa} \pdv{\bmkappa}{x}.
%    &= \pdv{E}{x}
%    = \bra{\Psi} \pdv{\hat{H}}{x} \ket{\Psi}.
\end{split}
\end{equation}
From the CI and Orbital stationarity conditions
$\partial E / \partial \bfc = 0 $ and $\partial E / \partial \bmkappa = 0$ (fulfilled within such fully variational methods), the nuclear gradient reduces to
the expectation value of the Hamiltonian derivative $\partial \hat{H} / \partial x$ (see Refs.~\cite{helgaker1984second,helgaker1998gradient,staalring2001analytical,yalouz2022analytical} and Supplementary Material~\ref{supp:nuclear_gradient})
\begin{equation}\label{eq:simplified_analytical_form_2}
\begin{split}
    \dv{E}{x} &= \pdv{E}{x}
%    + \pdv{E}{\bfc} \pdv{\bfc}{x}
%    + \pdv{E}{\bmkappa} \pdv{\bmkappa}{x} \\
%    &= \pdv{E}{x}
    = \bra{\Psi} \pdv{\hat{H}}{x} \ket{\Psi}.
\end{split}
\end{equation}  
As shown in Eq.~(\ref{eq:stationarity}), the OC-CASSCF approach fulfils similar stationarity conditions on both CI and orbital parameters. 
%As a result,
Consequently, the OC-CASSCF analytical gradients are 
% \vincent{As a result...therefore... : j'ai simplifié pour..."Consequently" !}
%therefore 
expected to retain the same form as in %Eq.~(\ref{eq:simplified_analytical_form}) 
Eq.~(\ref{eq:simplified_analytical_form_2}) 
for any optimized electronic state (ground and excited). 
In the following, we will show that this approach indeed yields consistent results compared to a numerical finite-difference approach.

\begin{figure*}[t]
    \centering
    \includegraphics[width=18cm]{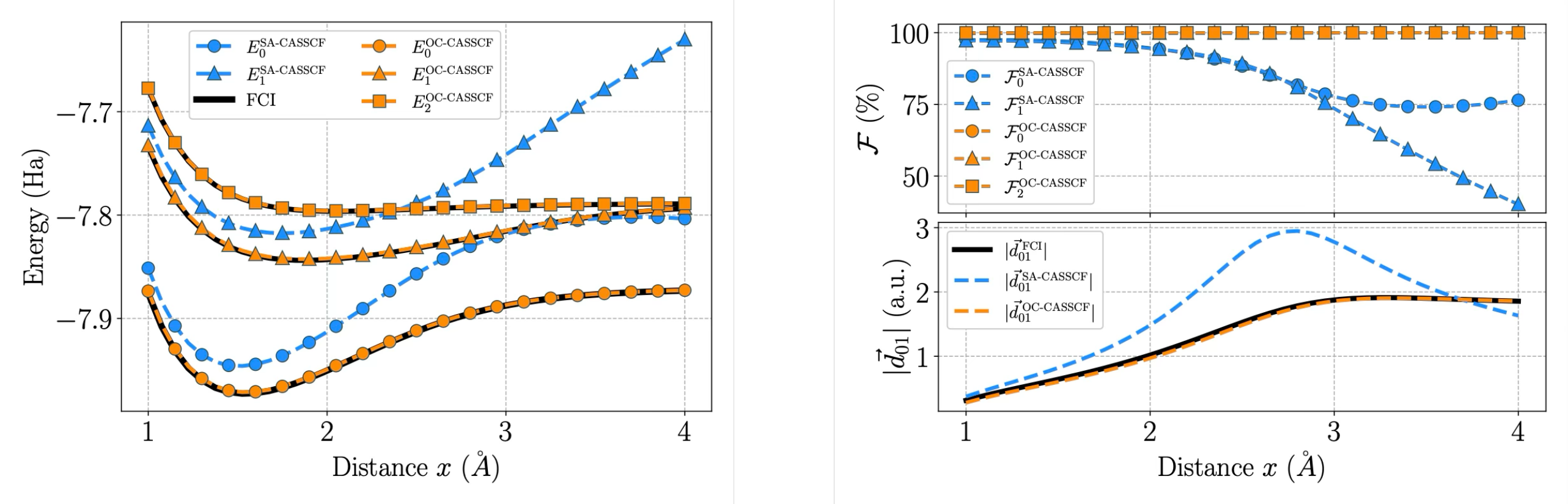} 
    \caption{
{Evolution of energies and wavefunction-based properties for LiH (with a minimal STO-6G basis set and a penalty term $\Delta = 1$~Ha). 
Left panel: Dissociation curves for the three low-lying singlet states of the molecule.}
The solid black curves show the FCI reference results. 
The blue curves with markers represent the ground and first excited states obtained using the SA-CASSCF method.
Similarly, the orange curves with markers correspond to the ground, first and second excited states obtained with the OC-CASSCF method.
Right top panel:  evolution of the state fidelity (as defined in Eq.~(\ref{eq:Fidelity})) of the SA-CASSCF and OC-CASSCF states relative to the FCI reference.
Right bottom panel: evolution of the transition dipole magnitude betwene the ground and first excited states (as defined in Eq.~(\ref{eq:trans_di})) obtained with  SA-, OC-CASSCF and FCI.
% \vincent{unité de $\Delta$ ? Hartree ?}.
% Highest energy error OC-CASSCF 2 e-3 Ha \\
% Highest energy error SA-CASSCF 1.6 e-1 Ha \\
% Lowest fidelity OC-CASSCF at most 99.72\% \\
% Fidelities SA-CASSCF   74 , 40\\
% Error dipole OC-CASSCF at most $5 \times 10^{-2}$ a.u.\\
% Error dipole SA-CASSCF at most $1.2$ a.u.
    }
    \label{fig:LiH_Energy_WFT}
\end{figure*}

\section{Numerical results}
\label{sec:numerical}

We now illustrate the performance of the OC-CASSCF method on the \chemfig{LiH} molecule (see Supplementary Material~\ref{supp:H2O} for similar
inspections  on the \chemfig{H_2O} molecule).
We focus on the three lowest singlet states and compare the results with 
%Full Configuration Interaction
FCI references, as well as with standard SA-CASSCF calculations.
% \vincent{"For each system..." est peut être de trop puisque seul LiH apparait DANS le main text. On pourrait même supprimer la phrase complète.}
% For each system, special attention is paid to energies, wavefunctions, and response properties.
All results and method implementations are obtained using the open-source package \textit{QuantNBody}~\cite{yalouz2022quantnbody}, which enables a systematic representation of second-quantized operators and many-body wavefunctions.
This package is used in conjunction with the quantum chemistry software \textit{Psi4}~\cite{smith2020psi4}, which provides the required \textit{ab initio} quantities.
% \vincent{J'arrêterais la phrase ici. Je l'ai fait}
%, such as electronic integrals.
All calculations are performed using a minimal STO-6G basis set, without imposing any spatial symmetry.

\subsection{Energies and Wavefunctions Properties}

% Let us first concentrate on  the energies and wavefunction properties.
Energies and wavefunction properties for LiH 
% on  the energies and wavefunction properties
are shown 
on Fig.~\ref{fig:LiH_Energy_WFT} as a function of the internuclear distance $x$.
FCI references account for the correlation of  
four electrons across six orbitals, whereas both the OC-CASSCF and SA-CASSCF calculations were performed at the CAS(2,2) level.
One should mention that OC-CASSCF calculations were run to describe all three low-lying states of the system (as in FCI), while SA-CASSCF were restricted to the ground and first excited states.
% \vincent{J'ai un peu changé la phrase précédente.}
%two low-lying states.

In the left panel of Fig.~\ref{fig:LiH_Energy_WFT}, the 
dissociation curves
%potential energy surfaces (PES) 
% \vincent{Pas vraiment une "surface", puisque une seule coordonnée. Parfois on parle de "dissociation curve". J'ai également fait cette proposition dans la légende de la Figure 1.}
obtained with OC-CASSCF are shown as orange dashed lines, SA-CASSCF as blue dashed lines, and FCI references as solid black lines.  
As readily seen, the three low-lying OC-CASSCF energies closely match the FCI references across the entire range of internuclear distances.  
Additional numerical inspections (not shown in Fig.~\ref{fig:LiH_Energy_WFT}) 
% \vincent{C'est un résultat ici. Je formulerais au présent, et non pas au passé}
%indicated
indicate here a maximum energy error of  $\sim 2\times 10^{-3}$~Ha with respect to FCI.  
% \vincent{Je propose de faire une phrase des deux suivantes. La proposition suit.}
% In contrast, the SA-CASSCF energies deteriorate rapidly as the distance increases.
% and completely diverge for $x > 1.5~\text{\AA}$. 
% In this second case, the maximum energy error relative to FCI for the two SA-CASSCF states is $\sim 0.16$~Ha, which is more considerable. 
% \vincent{SUGGESTION : In contrast, the SA-CASSCF energies deteriorate rapidly as the distance increases, 
% exhibiting a much stronger $\sim 0.16$~Ha deviation.
% and completely diverge for $x > 1.5~\text{\AA}$. 
{In contrast, the SA-CASSCF energies deteriorate rapidly as the distance increases, 
exhibiting a much stronger $\sim 0.16$~Ha deviation.
}
%\vincent{REMARQUE dont on doit reparler ! Sur ce point, on pourra nous rétorquer que nous nous limitons à CASSCF, et que PT2 pourrait nous ramener dans les clous de FCI. Nous pourrions laisser  penser avec la démarche que nous imaginons tout récupérer, systématiquement, avec un CASSCF. Evidemment, le point de départ OC-CASSCF est bien meilleur, à tel point qu'avec un CAS minimal, on retrouve FCI pour le fondamental (ce qui rarement le cas) et les excités. Il est vrai que les méthodes pertrubatives convergent lentement, voire mal. D'autre part, la reproduction de l'énergie FCI n'est pas un objectif, les différences font la spectroscopie (exemple : la méthode DDCI qui ne calcule qu'une petite partie de l'énergie de corrélation. Donc chaque état est très loin du FCI mais la spectroscopie est établie avec une grande précision). \textbf{Je ne change pas le texte. Je préfère que l'on échange ensemble sur ce point}. Les propriétés sont parlantes, pour commencer les DIFFERENCES d'énergie. Nous pourrions donner les énergies de dissociation E(x=infini) - E(xeq). Evidemment, la suite est TRES convaincante, avec la fidélité et les moments de transition. Peut être pourrait-on se rapprocher de nos autres activités récentes RSBW ?}

%Remarkably, 
% \vincent{J'ai commenté "Remarkably"}
Even with a
% \vincent{Il est plus que "modest", il est "minimal" !}
%modest
minimal
CAS(2,2) active space, these results reveal that OC-CASSCF provides a more accurate description of the three low-lying energies as compared to SA-CASSCF, which optimizes only two states (same observations as in Refs.\cite{saade2024excited,marie2023excited}).
% Additional simulations (not shown here) indicate that including a third state within SA-CASSCF further deteriorates the energy spectrum.
% with the only way to improve accuracy being an increase in the active-space size.

% The only way to reduce discrepancies with FCI in this case is to increase the size of the active space.

\begin{figure*}[t]
    \centering
    \includegraphics[width=2\columnwidth]{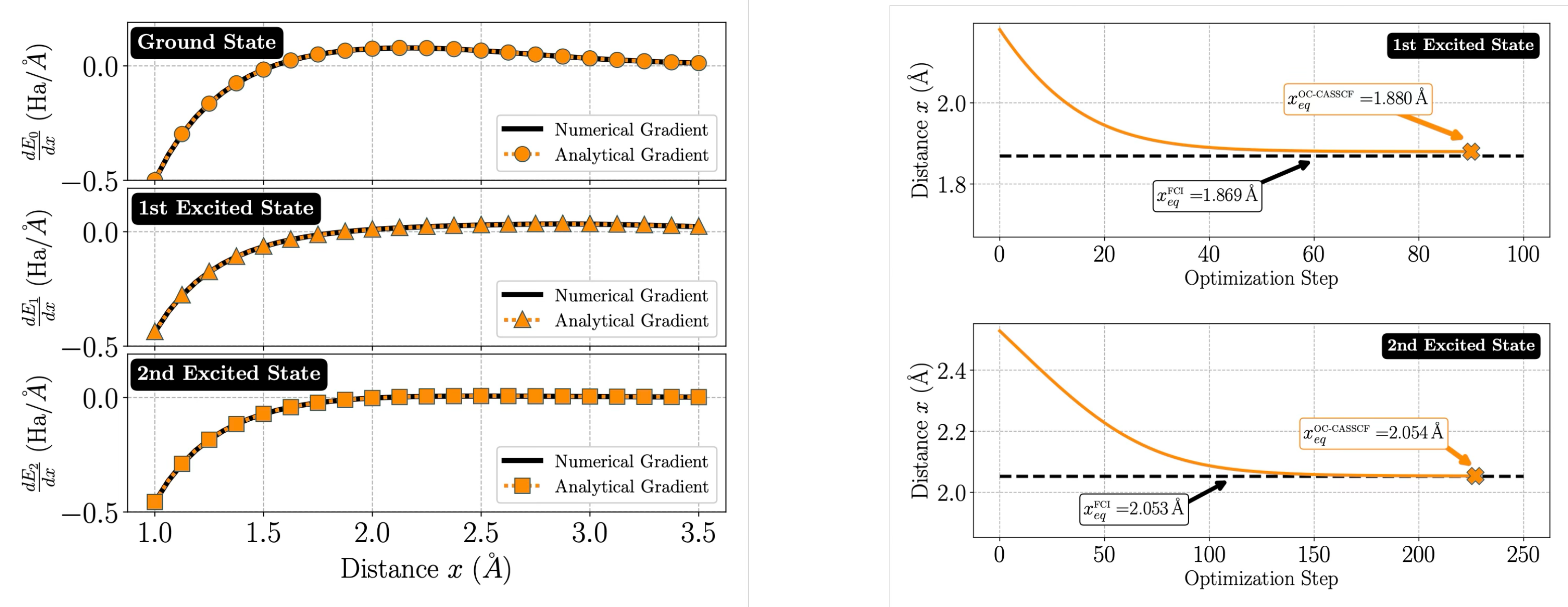} 
    \caption{ Illustration of the nuclear gradients computation for the LiH molecule (using  minimal STO-6G basis set with $\Delta = 1$~Ha).
    Left panels: Gradient amplitudes across the internuclear distance for the ground, first and second excited states.
    Analytical gradients are shown with orange curves, while black full lines are used for numerical gradients.
Right panels: Geometry optimization for the two excited states using a gradient descent approach with $\eta = 1$ (see Eq.~(\ref{eq:gradient_descent})).
Evolution of internuclear distance $x$ at each optimization step (orange full line).
FCI equilibrium positions are shown with black dashed lines.
% and the SA-CASSCF equilibrium position (blue dashed line). 
% All calculations were performed using a minimal STO-6G basis set with $\Delta = 1$~Ha and $\eta = 1$.
}
    \label{fig:LiH_gradient} 
\end{figure*}

%Beyond accurately reproducing energies, it is also essential for a method to provide reliable electronic wavefunctions and related properties.
% \vincent{J'ai un peu reformulé : To be reliable,  a method should not only provide accurate energy differences but 
% also proper electronic wavefunctions and related properties.}
To be reliable, a method should not only provide accurate energies but also proper electronic wavefunctions and related properties.
Therefore, we considered 
% \vincent{J'ai formulé au passé, "considered" plutôt que "consider".}
two complementary wavefunction-based measures.
The first one is the state-specific fidelity
% \vincent{C'est le carré du recouvrement. Peut-être le formuler ainsi ? J'ai proposé ci-après}
with respect to the FCI wavefunctions, 
% \vincent{i.e. the square of the overlap between the FCI and the method wavefunctions}:
%\vincent{Et j'ai mis un "m" minuscule du coup pour "method". Je ré-acualise ce soir samedi 17/01 ce commentaire. Il faudra harmoniser "m" ou "M". Il y en a d'autres plus loin.}
\begin{equation}\label{eq:Fidelity}
\mathcal{F}^\textrm{Method}_K = \left| \langle \Psi^\textrm{FCI}_K \mid \Psi^\textrm{Method}_K \rangle \right|^2 .
\end{equation}
% \vincent{Je ne donnerais pas la phrase suivante. Le terme
% "fidelity", la notion de "recouvrement".... : je pense que c'est très clair. Je supprimerais donc ceci : }
Following this definition, the closer 
$\mathcal{F}^\textrm{Method}_K$ 
%\vincent{m ou M ?}
is to $100\%$, the closer the wavefunction obtained with a given CAS "Method" (here, either OC- or SA-CASSCF) is to the exact FCI state.

In the top-right panel of Fig.~\ref{fig:LiH_Energy_WFT}, we report the evolution of the OC- and SA-CASSCF state fidelities as a function of the internuclear distance $x$.
% \vincent{$x$ est bien la distance}.
The OC-CASSCF results (orange curves) for the three electronic states remain very close to unity over the entire range of $x$ distance values.
% \vincent{J'ai commenté la fin de phrase, forme d'évidence.}
%, indicating excellent agreement with the FCI reference.
A more detailed analysis (not shown) reveals that the 
% \vincent{petit changement dans la formulation, pour désigner le type de fidélité (OC-CASSCF ici), et ensuite donner le minorant calculé (99.7)}
$\mathcal{F}^\textrm{OC\text{-}CASSCF}$ fidelity consistently exceeds $99.7\%$ for all three states.
In contrast, the fidelities obtained with SA-CASSCF deteriorate markedly as the internuclear distance $x$ increases.
% , particularly beyond $x > 1.5~\text{\AA}$.
In particular, the lowest fidelities 
% \vincent{Je reformule à peine}
%decrease to
can be as low as
$70\%$ and $40\%$ for the ground and first excited states, respectively.
% \vincent{Je supprime la fin de la phrase. On a démarré ce petit paragraphe par "deteriorate", et la on insiste encore. Je crois que le message est très convaincant et clair.}
%, indicating a significantly poorer representation of the exact FCI 
% wavefunctions as compared to OC-CASSCF states.

As a second measure of the wavefunction quality, we consider a physically motivated observable: the transition dipole moment.
The latter provides direct insight into the electronic redistribution upon photo-excitation and plays a central role in  optical transition probabilities and spectroscopic intensities.
More specifically, we evaluate the magnitude of the transition dipole between the ground and first excited states 
\begin{equation}\label{eq:trans_di}
|\vec{d}_{01}^\textrm{ Method}| = \sqrt{ \sum_{\alpha = X,Y,Z} \left| \left\langle \Psi^\textrm{Method}_0  | \hat{d}^{(\alpha)}  | \Psi^\textrm{Method}_1 \right\rangle \right|^2 },
\end{equation}
where $\hat{d}^{(\alpha)}$ is the dipole operator along the $\alpha$ direction 
%``$\alpha$'' 
%as defined by
\begin{equation}
\hat{d}^{(\alpha)} = \sum_{pq} \left( d^{\text{elec}(\alpha)}_{pq} + d^{\text{nuc} (\alpha)} \delta_{pq} \right) \hat{E}_{pq},
\end{equation}
where ${d}_{pq}^{\textrm{elec}\ (\alpha)}$ and ${d}^{\textrm{nuc} (\alpha)}$ are the electronic dipole integrals and the scalar nuclear dipole, respectively.
% \vincent{La phrase suivante est une répétition, voir sous  Eq.(13). On peut supprimer je pense.} 
 Note in Eq.~(\ref{eq:trans_di}) that the superscript "Method" 
%\vincent{m ou M ?}
 will cover the three approaches FCI, OC- and SA-CASSCF.
The resulting dipole magnitudes are shown in the bottom-right panel of Fig.~\ref{fig:LiH_Energy_WFT} as a function of the internuclear distance $x$.
% The exact FCI reference (black curve) exhibits a two-phase behavior: 
% \vincent{"...two-phase..." je n'adhère pas trop. SUGGESTION : }
The exact FCI reference (black curve) exhibits two distinct behaviors:
a nearly monotonic increase for $x \in [1\text{\AA}, 3\text{\AA}]$, followed by a \textit{ca.} $2$~a.u. plateau.
%at approximately $2$~a.u. 
% \vincent{Foire à la culottes avec les unités. Il faut décider : a.u. ou Angstrom. Quelle andouille je suis ! Je n'avais pas compris  la phrase. C'est la valeur du moment de transition !}
% beyond.
%for larger distances.
The OC-CASSCF results (orange curve) closely reproduce this trend, providing further evidence of the high quality of the ground and first excited wavefunctions.
% \vincent{J'ai coupé la fin de la phrase}
%obtained with this approach.
A more detailed numerical analysis (not shown here) indicates that the maximum deviation from the FCI reference remains below $5 \times 10^{-2}$~a.u. over the entire range of internuclear distances.
In contrast, the SA-CASSCF method significantly overestimates the transition dipole and %, reaching values as high as $3$~a.u., before decreasing sharply.
%at larger internuclear distances.
additional numerical analysis indicate a maximum deviation of $1.6$~a.u. compared to  FCI results. 
 
Overall, these observations, together with the state fidelity and the energy comparisons with respect to FCI, demonstrate that OC-CASSCF provides a more consistent description of both electronic energies and wavefunction-related properties than SA-CASSCF. 
Taken together, these results provide clear evidence that the use of a shared set of optimized orbitals within a state-averaged approach can sometimes significantly compromise the quality of both energies and wavefunctions.
Note that similar observations were also obtained for the H$_2$O molecule, as reported in Supplementary Material~\ref{supp:H2O}.

%\vincent{17 janvier : parfait pour moi. Il reste à échanger entre nous sur la manière d'amener les choses (voir page 4, commentaires que j'ai laissés). CASSCF n'est pas la fin de l'histoire. De pousser pour dire qu'en améliorant la description CAS on s'assure un traitement PT2 moins chaotique. Et peut être un écho aux papiers et développements RSBW ?}

\subsection{Nuclear Gradients and Geometry Optimizations}

To conclude on the LiH system, we discuss the evaluation of nuclear gradients within the OC-CASSCF framework and illustrate their use in a practical geometry optimization.
%: the geometry optimization of excited states.
%
We first examine the three left panels of Fig.~\ref{fig:LiH_gradient}, which show the evolution of gradient amplitudes for the ground, first and second excited states as a function of the internuclear distance $x$.
Here, analytical gradient amplitudes computed from Eq.~(\ref{eq:simplified_analytical_form_2}) (orange curves) are compared with numerical gradients (black curves) obtained \textit{via} a two-point finite-difference scheme.
For all three states, excellent agreement is observed between the analytical and numerical approaches, with a negligible discrepancy of at most $10^{-4}$~Ha/\AA\ across the full range of internuclear distances $x$.
%which thus confirm the predictive power of the analytical approach.
% \vincent{La fin phrase est troublante : on prouve l'approche analytique par une approche numérique. Je m'attendrais plutôt à l'inverse.
% Sauf que Newton-Raphson donc premier ordre, ou autre chose qui m'échappe ?}
% \saad{ Je comprends ta reflexion, et je pense que nous pouvons simplifier cette phrase en posant simplement la discrepency sans en faire trop sur le cote "predictive power".  J'ai commenté ! }

% \vincent{La phrase qui suit est une anaphore "not shown" du texte. Je pense qu'on pourrait en faire l'économie
% et donner ce chiffre dans la phrase précédente. Et je préfère parler de "negligible discrepancy", plutôt que de "visusally indinstiguishable" qui est moins convaincant. J'ai apporté ces changements dans la phrase qui précède, et j'ai commenté la phrase qui suivait. }
%A closer numerical inspection (not shown) reveals only a negligible discrepancy of at most $\sim 10^{-4}$~Ha/\AA\ between the two methods.

% Consistent with the  shapes of the dissociation curves in Fig.~\ref{fig:LiH_Energy_WFT}, each gradient crosses zero only once, confirming the existence of a single energy minimum for each state.
% \vincent{Ce point est apparent sur l'état fondamental. Pour le premier et second excités, c'est moins visible. On pourrait zoomer en traçant les courbes dans l'intervalle 1.25 et 3 sur la Figure 2. Bien sûr l'intervalle ne serait plus le même que celui de la Figure 1.}

As a practical application, we employed the analytical gradients to perform geometry optimizations of the two singlet excited states of LiH.
As previously shown in Fig.~\ref{fig:LiH_Energy_WFT}, the FCI dissociation curves for the first and second excited states each display a single energy minimum, with equilibrium distances of $x_{eq}^{\mathrm{FCI}} = 1.87\text{\AA}$ and $2.05\text{\AA}$, respectively.
% Motivated by these observations, we employed the analytical gradients to perform geometry optimizations of the two excited states. 
To compare, the equilibrium distances $x^\textrm{OC-CASSCF}_{eq}$ that define the OC-CASSCF  
dissociation curves
%PES 
minima were located \textit{via} a gradient-descent procedure:
\begin{equation}\label{eq:gradient_descent}
x^{n+1} = x^{n} - \eta \left.\frac{dE_K}{dx}\right|_{x = x^{n}},
\end{equation}
where $\eta$ is a fixed damping parameter,
% \vincent{On donne une valeur de $\eta$ ? J'ai le souvenir de travailler avec ces dampings, mais plus de souvenir}
$x^{n}$ is the geometry at step $n$, and $\frac{dE_K}{dx}$ is evaluated based on Eq.~(\ref{eq:simplified_analytical_form_2}).
The top-right and bottom-right panels of Fig.~\ref{fig:LiH_gradient} show the optimization trajectories for the first and second excited states, respectively (see caption for simulation details).
Interestingly, for both states, the optimized OC-CASSCF geometries are very close to the FCI equilibrium points, 
% \vincent{Je combine les deux idées de précision et de chiffrage de l'écart en une inégalité.}
 with $\vert x^\textrm{OC-CASSCF}_{eq} \sim x^\textrm{FCI}_{eq} \vert < 10^{-2}$~\AA.
%$x^\textrm{OC-CASSCF}_{eq} \sim x^\textrm{FCI}_{eq}$, with a maximum discrepancy of only $10^{-2}$~\AA.  

These results further underscore the high quality of the dissociation curves
%PES
produced by OC-CASSCF compared to FCI, and also confirm the accuracy and reliability of the analytical gradient in practical geometry optimizations for locating the associated spectral features.
Note again that similar observations were also obtained for the H$_2$O molecule, as reported in Supplementary Material~\ref{supp:H2O}.

% \textcolor{blue}{
% Direct comparisons with FCI are feasible here only due to the small molecular sizes and associated basis sets. For larger systems, such comparisons are generally impractical, and CAS calculations are typically combined with perturbation theory to approach FCI accuracy. In this context, the good agreement between OC-CASSCF and FCI observed for small systems suggests that OC-CASSCF can provide a reliable zeroth-order reference for subsequent perturbative treatments~\cite{roos1982simple,andersson1992second,angeli2001introduction,angeli2004quasidegenerate} in larger systems.
% }

\section*{Conclusion}

In this work, we have presented several theoretical extensions to the previously introduced orthogonally constrained version of the CASSCF method (OC-CASSCF)~\cite{yalouz2023ocoo}. 
We introduced a two-step procedure based on a second-order Newton-Raphson scheme for orthogonally constrained orbital optimization, for which analytical expressions of the constrained orbital Gradients and Hessians have been derived. 
As a possible extension toward a direct one-step formulation, complementary Gradients and Hessians for the CI and orbital-CI blocks are also provided (see Supplementary Material~\ref{supp:configuration_opt}). 
In addition, we have discussed a route to compute analytical nuclear gradients, enabling the application of OC-CASSCF to geometry optimizations and related studies.
Based on these tools, the performance of the OC-CASSCF procedure was assessed by optimizing the three lowest singlet states of \chemfig{LiH} (and \chemfig{H_2O} in Supplementary material \ref{supp:H2O}).
The resulting energies and states show very good agreement with reference Full CI results, whereas SA-CASSCF fails to provide an adequate description.
%\vincent{Revenant sur le point déjà discuté plus haut. Je pense que nous devons insister en argumentant sur l'intérêt d'avoir un meilleur CAS pour éventuellement aller au delà, typiquement CASPT2. "SA-CASSCF + PT2" (j'ai l'expérience des interactions d'échange dans Ising ou Heisenberg) a connu beaucoup de succès, sans pour autant que SA-CASSCF soit de qualité spectroscopique. Et de  poursuivre la croisade "chaque état dans son jeu d'OM". \textbf{Je ne change pas le texte. Je préfère que l'on échange ensemble sur ce point}.}
% The analytical nuclear gradients were further employed in geometry optimizations to locate the local minima of the first and second excited states of these molecules, yielding equilibrium geometries consistent with Full CI references.
% \vincent{Plutot que ce qui précède, je propose pour éviter la répétition de la confrontation a full CI.
{Analytical nuclear gradients were further employed in geometry optimizations to accurately locate the local minima of the first and second excited states.}

Naturally, this work should be regarded as preliminary, focusing on the prototyping of OC-CASSCF for small molecules using minimal basis sets. 
The very good agreement observed with FCI results is encouraging, but should not be considered definitive or systematic. 
Larger-scale simulations of more complex systems will be required in future, and we hope that the tools and demonstrations presented here  will facilitate future implementations of OC-CASSCF in advanced electronic-structure packages. 
For larger and more practical systems, CAS calculations will naturally require subsequent perturbative treatments (e.g., CASPT2 or NEVPT2~\cite{roos1982simple,andersson1992second,angeli2001introduction,angeli2004quasidegenerate}) to approach FCI accuracy.
In this broader context, it remains to be explored whether OC-CASSCF can provide a more reliable zeroth-order description than the standard state-averaged approach for such perturbative corrections.
These questions are left for future work.

% Building on the theoretical developments and numerical demonstrations presented above, we expect this work to motivate and facilitate future implementations in higher-level electronic structure packages, enabling the assessment of the method on larger molecular systems.
% Naturally, the present work focuses on prototyping the OC-CASSCF method, with small molecules and minimal basis sets.  
% Within this context, it would be interesting to explore how OC-CASSCF can provide a more reliable zeroth-order description than the state-averaged approach for subsequent perturbative treatments. 
% These questions are left for future work. 
% This, in turn, will enable systematic assessments on larger molecular systems and active spaces, as well as the study of complex excited state phenomena.

% Direct comparisons with FCI are feasible here only due to the small molecular sizes and associated basis sets. For larger systems, such comparisons are generally impractical, and CAS calculations are typically combined with perturbation theory to approach FCI accuracy. In this context, the good agreement between OC-CASSCF and FCI observed for small systems suggests that OC-CASSCF can provide a reliable zeroth-order reference for subsequent perturbative treatments~\cite{roos1982simple,andersson1992second,angeli2001introduction,angeli2004quasidegenerate} in larger systems.

\section*{Supplementary Material}

\begin{enumerate}
    \item \label{supp:orbital_opt} Derivation of the analytical expressions for the overlap orbital Gradient and Hessian.

    \item \label{supp:cas_trdm} Analytical expressions for the TRDMs when one of the states belongs to the CAS.

    \item \label{supp:configuration_opt} Derivation of analytical expressions for the overlap configuration Gradient and Hessian and for the overlap mixed Hessian in the case of a one-step optimization.

    \item \label{supp:nuclear_gradient} Analytical expression of the derivative of the Hamiltonian operator.

    \item \label{supp:H2O} Numerical results for the first and second excited states of the \chemfig{H_2O} molecule (potential energy surfaces, transition dipole moments and geometry optimization).
\end{enumerate}

\begin{acknowledgements}
    This work benefited from State support managed by the ANR under the France 2030 program, referenced by ANR-23-PETQ-0006. 
    It is also supported by the Interdisciplinary Thematic Institute QMat, as part of the ITI 2021-2028 program of the University of Strasbourg, CNRS and Inserm, and was supported by IdEx Unistra (ANR-10-IDEX-0002), SFRI STRAT’ US project (ANR-20-SFRI-0012), and EUR QMAT (ANR-17-EURE-0024). Both these fundings were granted under the framework of the French Investments for the Future Program.
\end{acknowledgements}

% \nocite{*}
\bibliography{biblio.bib}% Produces the bibliography via BibTeX.

@preamble{"\newcommand{\Aa}[0]{Aa} " }

@STRING{IJQC="Int. J. Quantum Chem."}

@STRING{JCP="J. Chem. Phys."}

@STRING{MP="Mol. Phys."}

@STRING{PR="Phys. Rev."}

@STRING{QST="Quantum Sci. Technol."}

@STRING{RA="Radiochim. Acta"}

@article{andersson1992second,
  title={Second-order perturbation theory with a complete active space self-consistent field reference function},
  author={Andersson, Kerstin and Malmqvist, Per-{\AA}ke and Roos, Bj{\"o}rn O},
  journal=JCP,
  volume={96},
  number={2},
  pages={1218--1226},
  year={1992},
  publisher={American Institute of Physics},
  url={https://doi.org/10.1063/1.462209}
}

@article{angeli2001introduction,
  title={Introduction of n-electron valence states for multireference perturbation theory},
  author={Angeli, Celestino and Cimiraglia, Renzo and Evangelisti, S and Leininger, T and Malrieu, J-P},
  journal=JCP,
  volume={114},
  number={23},
  pages={10252--10264},
  year={2001},
  publisher={American Institute of Physics},
  url={https://doi.org/10.1063/1.1361246}
}

@article{higgott2019variational,
  title={Variational quantum computation of excited states},
  author={Higgott, Oscar and Wang, Daochen and Brierley, Stephen},
  journal={Quantum},
  volume={3},
  pages={156},
  year={2019},
  publisher={Verein zur F{\"o}rderung des Open Access Publizierens in den Quantenwissenschaften},
  url={https://doi.org/10.22331/q-2019-07-01-156}
}

@article{nakanishi2019subspace,
  title={Subspace-search variational quantum eigensolver for excited states},
  author={Nakanishi, Ken M and Mitarai, Kosuke and Fujii, Keisuke},
  journal={Phys. Rev. Res.},
  volume={1},
  number={3},
  pages={033062},
  year={2019},
  publisher={APS},
  url={https://doi.org/10.1103/PhysRevResearch.1.033062}
}

@article{siegbahn1981complete,
  title={The complete active space SCF (CASSCF) method in a Newton--Raphson formulation with application to the HNO molecule},
  author={Siegbahn, Per EM and Alml{\"o}f, Jan and Heiberg, Anders and Roos, Bj{\"o}rn O},
  journal=JCP,
  volume={74},
  number={4},
  pages={2384--2396},
  year={1981},
  publisher={American Institute of Physics},
  url={https://doi.org/10.1063/1.441359}
}

@book{helgaker2014molecular,
  title={Molecular electronic-structure theory},
  author={Helgaker, Trygve and Jorgensen, Poul and Olsen, Jeppe},
  year={2014},
  publisher={John Wiley \& Sons}
}

@article{smith2020psi4,
  title={PSI4 1.4: Open-source software for high-throughput quantum chemistry},
  author={Smith, Daniel GA and Burns, Lori A and Simmonett, Andrew C and Parrish, Robert M and Schieber, Matthew C and Galvelis, Raimondas and Kraus, Peter and Kruse, Holger and Di Remigio, Roberto and Alenaizan, Asem and others},
  journal=JCP,
  volume={152},
  number={18},
  pages={184108},
  year={2020},
  publisher={AIP Publishing LLC},
  url={https://doi.org/10.1063/5.0006002}
}

@article{malmqvist2002restricted,
  title={The restricted active space (RAS) state interaction approach with spin--orbit coupling},
  author={Malmqvist, Per {\AA}ke and Roos, Bj{\"o}rn O and Schimmelpfennig, Bernd},
  journal={Chem. Phys. Lett.},
  volume={357},
  number={3-4},
  pages={230--240},
  year={2002},
  publisher={Elsevier},
  url={https://doi.org/10.1016/S0009-2614(02)00498-0}
}

@article{staalring2001analytical,
  title={Analytical gradients of a state average MCSCF state and a state average diagnostic},
  author={St{\aa}lring, Jonna and Bernhardsson, Anders and Lindh, Roland},
  journal=MP,
  volume={99},
  number={2},
  pages={103--114},
  year={2001},
  publisher={Taylor \& Francis},
  url={https://doi.org/10.1080/002689700110005642}
}

@article{helgaker1984second,
  title={A second-quantization approach to the analytical evaluation of response properties for perturbation-dependent basis sets},
  author={Helgaker, Trygve U and Alml{\"o}f, Jan},
  journal=IJQC,
  volume={26},
  number={2},
  pages={275--291},
  year={1984},
  publisher={Wiley Online Library},
  url={https://onlinelibrary.wiley.com/doi/pdf/10.1002/qua.560260211?casa_token=9m1l-5msg9kAAAAA:tMp9pdPEXayc-miP7cyPQ3scQT1xFe6zVgtYlmQrGbZF_aWtL1F89q2gqNbUlMpYFDZf-Ni2zaitu6Pj}
}

@article{Yalouz_2021,
	doi = {10.1088/2058-9565/abd334},
	url = {https://doi.org/10.1088/2058-9565/abd334},
	year = 2021,
	volume = {6},
	number = {2},
	pages = {024004},
	author = {Saad Yalouz and Bruno Senjean and Jakob Günther and Francesco Buda and Thomas E O'Brien and Lucas Visscher},
	title = {A state-averaged orbital-optimized hybrid quantum{\textendash}classical algorithm for a democratic description of ground and excited states},
	journal = QST
}

@article{yalouz2023ocoo,
    author = {Yalouz, Saad and Robert, Vincent},
    year = {2023},
    month = {02},
    pages = {1381-1640},
    title = {Orthogonally Constrained Orbital Optimization: Assessing Changes of Optimal Orbitals for Orthogonal Multireference States},
    volume = {19},
    issue = {5},
    journal = {Journal of chemical theory and computation},
    doi = {10.1021/acs.jctc.2c01144}
}

@article{angeli2004quasidegenerate,
  title={A quasidegenerate formulation of the second order n-electron valence state perturbation theory approach},
  author={Angeli, Celestino and Borini, Stefano and Cestari, Mirko and Cimiraglia, Renzo},
  journal=JCP,
  volume={121},
  number={9},
  pages={4043--4049},
  year={2004},
  publisher={American Institute of Physics},
  url={https://doi.org/10.1063/1.1778711}
}

@misc{zhu2025statespecificoo,
      title={State-Specific Orbital Optimization for Enhanced Excited-States Calculation on Quantum Computers}, 
      author={Guorui Zhu and Joel Bierman and Jianfeng Lu and Yingzhou Li},
      year={2025},
      eprint={2510.13544},
      archivePrefix={arXiv},
      primaryClass={quant-ph},
      doi={abs/2510.13544}, 
}

@article{hunt1969orthogonality,
  title={The orthogonality constrained basis set expansion method for treating off-diagonal lagrange multipliers in calculations of electronic wave functions},
  author={Hunt, WJ and Dunning Jr, TH and Goddard III, WA},
  journal={Chemical Physics Letters},
  volume={3},
  number={8},
  pages={606--610},
  year={1969},
  publisher={Elsevier},
  url={https://doi.org/10.1016/0009-2614(69)85122-5}
}

@article{gilbert2008self,
  title={Self-consistent field calculations of excited states using the maximum overlap method (MOM)},
  author={Gilbert, Andrew TB and Besley, Nicholas A and Gill, Peter MW},
  journal={The Journal of Physical Chemistry A},
  volume={112},
  number={50},
  pages={13164--13171},
  year={2008},
  publisher={ACS Publications},
  url={https://doi.org/10.1021/jp801738f}
}

@article{barca2018simple,
  title={Simple models for difficult electronic excitations},
  author={Barca, Giuseppe MJ and Gilbert, Andrew TB and Gill, Peter MW},
  journal={Journal of chemical theory and computation},
  volume={14},
  number={3},
  pages={1501--1509},
  year={2018},
  publisher={ACS Publications},
  url={https://doi.org/10.1021/acs.jctc.7b00994}
}

@article{dash2021tailoring,
  title={Tailoring CIPSI expansions for QMC calculations of electronic excitations: the case study of thiophene},
  author={Dash, Monika and Moroni, Saverio and Filippi, Claudia and Scemama, Anthony},
  journal={Journal of chemical theory and computation},
  volume={17},
  number={6},
  pages={3426--3434},
  year={2021},
  publisher={ACS Publications},
  url={https://doi.org/10.1021/acs.jctc.1c00212}
}

@article{cuzzocrea2022reference,
  title={Reference excitation energies of increasingly large molecules: a QMC study of cyanine dyes},
  author={Cuzzocrea, Alice and Moroni, Saverio and Scemama, Anthony and Filippi, Claudia},
  journal={Journal of chemical theory and computation},
  volume={18},
  number={2},
  pages={1089--1095},
  year={2022},
  publisher={ACS Publications},
  url={https://doi.org/10.1021/acs.jctc.1c01162}
}

@article{yalouz2022quantnbody, 
doi = {10.21105/joss.04759}, 
url = {https://doi.org/10.21105/joss.04759},  
year = {2022}, 
publisher = {The Open Journal}, 
volume = {7}, 
number = {80}, 
pages = {4759}, 
author = {Yalouz, Saad and Gullin, Martin Rafael and Sekaran, Sajanthan}, 
title = {QuantNBody: a Python package for quantum chemistry and physics to build and manipulate many-body operators and wave functions.}, 
journal = {Journal of Open Source Software} }

@misc{helgaker1998gradient,
  title={Gradient theory},
  author={Helgaker, Trygve and Allinger, NL and Clark, T and Gasteiger, J and Kollmann, PA and SchaeferIII, HF and Schreiner, PR and others},
  journal={The Encyclopedia of Computational Chemistry},
  pages={1157--1169},
  year={1998},
  publisher={Wiley Chichester}
}

@article{yamaguchi2011analytic,
  title={Analytic derivative methods in molecular electronic structure theory: A new dimension to quantum chemistry and its applications to spectroscopy},
  author={Yamaguchi, Yukio and Schaefer III, Henry F},
  journal={Handbook of High-resolution Spectroscopy},
  year={2011},
  publisher={Wiley Online Library},
url={https://doi.org/10.1002/9780470749593.hrs006}
}

@article{yalouz2022analytical,
  title={Analytical nonadiabatic couplings and gradients within the state-averaged orbital-optimized variational quantum eigensolver},
  author={Yalouz, Saad and Koridon, Emiel and Senjean, Bruno and Lasorne, Benjamin and Buda, Francesco and Visscher, Lucas},
  journal={Journal of chemical theory and computation},
  volume={18},
  number={2},
  pages={776--794},
  year={2022},
  publisher={ACS Publications},
url={https://doi.org/10.1021/acs.jctc.1c00995}
}

@article{roseiro2025interplay,
  title={Interplay between Spinmerism and Spin-Orbit Coupling for a d2 Metal Ion in an Open-Shell Ligand Field},
  author={Roseiro, Pablo and Shah, Ashini and Yalouz, Saad and Robert, Vincent},
  journal={ChemPhysChem},
  volume={26},
  number={6},
  pages={e202400914},
  year={2025},
  publisher={Wiley Online Library},
url={https://doi.org/10.1002/cphc.202400914}
}

@article{roseiro2023modifications,
  title={Modifications of Tanabe-Sugano d 6 diagram induced by radical ligand field: ab initio inspection of a Fe (II)-verdazyl molecular complex},
  author={Roseiro, Pablo and Yalouz, Saad and Brook, David JR and Ben Amor, Nadia and Robert, Vincent},
  journal={Inorganic Chemistry},
  volume={62},
  number={14},
  pages={5737--5743},
  year={2023},
  publisher={ACS Publications},
url={https://doi.org/10.1021/acs.inorgchem.3c00275}
}

@article{sheng1994magnetism,
  title={Magnetism and pairing in a C 60 molecule: A variational Monte Carlo study},
  author={Sheng, DN and Weng, ZY and Ting, CS and Dong, JM},
  journal={Physical Review B},
  volume={49},
  number={6},
  pages={4279},
  year={1994},
  publisher={APS},
url={https://doi.org/10.1103/PhysRevB.49.4279}
}

@article{verot2012importance,
  title={Importance of a multiconfigurational description for molecular junctions},
  author={V{\'e}rot, Martin and Borshch, Serguei A and Robert, Vincent},
  journal={Chemical Physics Letters},
  volume={519},
  pages={125--129},
  year={2012},
  publisher={Elsevier},
url={https://doi.org/10.1016/j.cplett.2011.11.013}
}

@article{vela2017electron,
  title={Electron transport through a spin crossover junction. Perspectives from a wavefunction-based approach},
  author={Vela, Sergi and Verot, Martin and Fromager, Emmanuel and Robert, Vincent},
  journal={The Journal of Chemical Physics},
  volume={146},
  number={6},
  year={2017},
  publisher={AIP Publishing},
url={https://doi.org/10.1063/1.4975327}
}

@article{thom2008locating,
  title={Locating multiple self-consistent field solutions: an approach inspired by metadynamics},
  author={Thom, Alex JW and Head-Gordon, Martin},
  journal={Physical review letters},
  volume={101},
  number={19},
  pages={193001},
  year={2008},
  publisher={APS},
url={http://dx.doi.org/10.1103/PhysRevLett.101.193001}
}

@article{pathak2021excited,
  title={Excited states in variational Monte Carlo using a penalty method},
  author={Pathak, Shivesh and Busemeyer, Brian and Rodrigues, Jo{\~a}o NB and Wagner, Lucas K},
  journal={The Journal of Chemical Physics},
  volume={154},
  number={3},
  year={2021},
  publisher={AIP Publishing},
url={https://doi.org/10.1063/5.0030949}
}

@article{stoudenmire2012studying,
  title={Studying two-dimensional systems with the density matrix renormalization group},
  author={Stoudenmire, Edwin M and White, Steven R},
  journal={Annu. Rev. Condens. Matter Phys.},
  volume={3},
  number={1},
  pages={111--128},
  year={2012},
  publisher={Annual Reviews},
url={https://doi.org/10.1146/annurev-conmatphys-020911-125018}
}

@article{shepard2022double,
  title={Double excitation energies from quantum Monte Carlo using state-specific energy optimization},
  author={Shepard, Stuart and Panades-Barrueta, Ramon L and Moroni, Saverio and Scemama, Anthony and Filippi, Claudia},
  journal={Journal of chemical theory and computation},
  volume={18},
  number={11},
  pages={6722--6731},
  year={2022},
  publisher={ACS Publications},
url={https://doi.org/10.1021/acs.jctc.2c00769}
}

@article{dash2019excited,
  title={Excited states with selected configuration interaction-quantum Monte Carlo: Chemically accurate excitation energies and geometries},
  author={Dash, Monika and Feldt, Jonas and Moroni, Saverio and Scemama, Anthony and Filippi, Claudia},
  journal={Journal of chemical theory and computation},
  volume={15},
  number={9},
  pages={4896--4906},
  year={2019},
  publisher={ACS Publications},
url={https://doi.org/10.1021/acs.jctc.9b00476}
}

@article{kossoski2021excited,
  title={Excited states from state-specific orbital-optimized pair coupled cluster},
  author={Kossoski, F{\'a}bris and Marie, Antoine and Scemama, Anthony and Caffarel, Michel and Loos, Pierre-Francois},
  journal={Journal of Chemical Theory and Computation},
  volume={17},
  number={8},
  pages={4756--4768},
  year={2021},
  publisher={ACS Publications},
url={https://doi.org/10.1021/acs.jctc.1c00348}
}

@article{damour2024state,
  title={State-specific coupled-cluster methods for excited states},
  author={Damour, Yann and Scemama, Anthony and Jacquemin, Denis and Kossoski, F{\'a}bris and Loos, Pierre-Fran{\c{c}}ois},
  journal={Journal of Chemical Theory and Computation},
  volume={20},
  number={10},
  pages={4129--4145},
  year={2024},
  publisher={ACS Publications},
url={https://doi.org/10.1021/acs.jctc.4c00034}
}

@article{marie2021variational,
  title={Variational coupled cluster for ground and excited states},
  author={Marie, Antoine and Kossoski, F{\'a}bris and Loos, Pierre-Fran{\c{c}}ois},
  journal={The Journal of Chemical Physics},
  volume={155},
  number={10},
  year={2021},
  publisher={AIP Publishing},
url={https://doi.org/10.1063/5.0060698}
}

@article{chan2016matrix,
  title={Matrix product operators, matrix product states, and ab initio density matrix renormalization group algorithms},
  author={Chan, Garnet Kin and Keselman, Anna and Nakatani, Naoki and Li, Zhendong and White, Steven R},
  journal={The Journal of chemical physics},
  volume={145},
  number={1},
  year={2016},
  publisher={AIP Publishing},
url={https://doi.org/10.1063/1.4955108}
}

@article{white1993density,
  title={Density-matrix algorithms for quantum renormalization groups},
  author={White, Steven R},
  journal={Physical review b},
  volume={48},
  number={14},
  pages={10345},
  year={1993},
  publisher={APS},
url={https://doi.org/10.1103/PhysRevB.48.10345}
}

@article{fishman2022itensor,
  title={The ITensor software library for tensor network calculations},
  author={Fishman, Matthew and White, Steven and Stoudenmire, Edwin Miles},
  journal={SciPost Physics Codebases},
  pages={004},
  year={2022},
url={url={https://scipost.org/10.21468/SciPostPhysCodeb.4}}
}

@article{dobrautz2021spin,
  title={Spin-pure stochastic-CASSCF via GUGA-FCIQMC applied to iron--sulfur clusters},
  author={Dobrautz, Werner and Weser, Oskar and Bogdanov, Nikolay A and Alavi, Ali and Li Manni, Giovanni},
  journal={Journal of Chemical Theory and Computation},
  volume={17},
  number={9},
  pages={5684--5703},
  year={2021},
  publisher={ACS Publications},
url={https://doi.org/10.1021/acs.jctc.1c00589}
}

@article{manni2023openmolcas,
  title={The OpenMolcas Web: A community-driven approach to advancing computational chemistry},
  author={Manni, Giovanni Li and Galv{\'a}n, Ignacio Fdez and Alavi, Ali and Aleotti, Flavia and Aquilante, Francesco and Autschbach, Jochen and Avagliano, Davide and Baiardi, Alberto and Bao, Jie J and Battaglia, Stefano and others},
  journal={Journal of chemical theory and computation},
  volume={19},
  number={20},
  pages={6933},
  year={2023},
url={https://doi.org/10.1021/acs.jctc.3c00182}
}

@article{marie2023excited,
  title={Excited states, symmetry breaking, and unphysical solutions in state-specific CASSCF theory},
  author={Marie, Antoine and Burton, Hugh GA},
  journal={The Journal of Physical Chemistry A},
  volume={127},
  number={20},
  pages={4538--4552},
  year={2023},
  publisher={ACS Publications},
url={https://doi.org/10.1021/acs.jpca.3c00603}
}

@article{saade2024excited,
  title={Excited state-specific CASSCF theory for the torsion of ethylene},
  author={Saade, Sandra and Burton, Hugh GA},
  journal={Journal of Chemical Theory and Computation},
  volume={20},
  number={12},
  pages={5105--5114},
  year={2024},
  publisher={ACS Publications},
url={https://doi.org/10.1021/acs.jctc.4c00212}
}

@article{bonnet2023ruthenium,
  title={Ruthenium-based photoactivated chemotherapy},
  author={Bonnet, Sylvestre},
  journal={Journal of the American Chemical Society},
  volume={145},
  number={43},
  pages={23397--23415},
  year={2023},
  publisher={ACS Publications},
url={https://doi.org/10.1021/jacs.3c01135}
}

@article{smith2013recent,
  title={Recent advances in singlet fission},
  author={Smith, Millicent B and Michl, Josef},
  journal={Annual review of physical chemistry},
  volume={64},
  number={1},
  pages={361--386},
  year={2013},
  publisher={Annual Reviews},
url={https://doi.org/10.1021/cr1002613}
}

@article{marcus1993electron,
  title={Electron transfer reactions in chemistry: theory and experiment (Nobel lecture)},
  author={Marcus, Rudolph A},
  journal={Angewandte Chemie International Edition in English},
  volume={32},
  number={8},
  pages={1111--1121},
  year={1993},
  publisher={Wiley Online Library},
url={https://doi.org/10.1002/anie.199311113}
}

@article{domingo2015electronic,
  title={Electronic reorganization triggered by electron transfer: The intervalence charge transfer of a Fe3+/Fe2+ bimetallic complex},
  author={Domingo, Alex and Angeli, Celestino and de Graaf, Coen and Robert, Vincent},
  journal={Journal of Computational Chemistry},
  volume={36},
  number={11},
  pages={861--869},
  year={2015},
  publisher={Wiley Online Library},
url={https://doi.org/10.1002/jcc.23871}
}

@article{levi2020variational,
  title={Variational calculations of excited states via direct optimization of the orbitals in DFT},
  author={Levi, Gianluca and Ivanov, Aleksei V and J{\'o}nsson, Hannes},
  journal={Faraday Discussions},
  volume={224},
  pages={448--466},
  year={2020},
  publisher={Royal Society of Chemistry},
url={https://doi.org/10.1039/D0FD00064G}
}

@article{roos1982simple,
  title={A simple method for the evaluation of the second-order-perturbation energy from external double-excitations with a CASSCF reference wavefunction},
  author={Roos, Bj{\"o}rn O and Linse, Per and Siegbahn, Per EM and Blomberg, Margareta RA},
  journal={Chemical Physics},
  volume={66},
  number={1-2},
  pages={197--207},
  year={1982},
  publisher={Elsevier},
url={https://doi.org/10.1016/0301-0104(82)88019-1}
}

\end{document}
%
% ****** End of file aipsamp.tex ******